\documentclass[aps,prb,preprint,superscriptaddress,showkeys,floatfix]{revtex4}

\usepackage{graphicx,color}
\graphicspath{{figs/}}

\begin{document}
\title{Enhancing the Mass Sensitivity of Graphene Nanoresonators Via Nonlinear Oscillations: The Effective Strain Mechanism}
\author{Jin-Wu Jiang}
    \affiliation{Institute of Structural Mechanics, Bauhaus-University Weimar, Marienstr. 15, D-99423 Weimar, Germany}
\author{Harold S. Park}
    \altaffiliation{Electronic address: parkhs@acs.bu.edu}
    \affiliation{Department of Mechanical Engineering, Boston University, Boston, Massachusetts 02215, USA}
\author{Timon Rabczuk}
    \altaffiliation{Electronic address: timon.rabczuk@uni-weimar.de}
    \affiliation{Institute of Structural Mechanics, Bauhaus-University Weimar, Marienstr. 15, D-99423 Weimar, Germany}

\date{\today}
\begin{abstract}
We perform classical molecular dynamics simulations to investigate the enhancement of the mass sensitivity and resonant frequency of graphene nanomechanical resonators that is achieved by driving them into the nonlinear oscillation regime.  The mass sensitivity as measured by the resonant frequency shift is found to triple if the actuation energy is about 2.5 times the initial kinetic energy of the nanoresonator.  The mechanism underlying the enhanced mass sensitivity is found to be the effective strain that is induced in the nanoresonator due to the nonlinear oscillations, where we obtain an analytic relationship between the induced effective strain and the actuation energy that is applied to the graphene nanoresonator.  An important implication of this work is that there is no need for experimentalists to apply tensile strain to the resonators before actuation in order to enhance the mass sensitivity.  Instead, enhanced mass sensitivity can be obtained by the far simpler technique of actuating nonlinear oscillations of an existing graphene nanoresonator.

\end{abstract}

\maketitle

\pagebreak

\section{introduction}

Carbon nanotube and graphene-based nanomechanical resonators (GNMR) have attracted significant attention from the scientific community; see the recent reviews of Refs.~\onlinecite{EkinciKL,ArlettJL,EomK,BartonRA}. Experiments have demonstrated that single-walled carbon nanotube nanomechanical resonators can serve as a mass sensor that is capable of detecting individual atoms or molecules~\cite{JensenK,LassagneB,ChiuHY}, which is made possible by the high stiffness and low mass of carbon nanotubes.  Graphene also possesses high stiffness~\cite{LeeCG,JiangJW2009,JiangJW2010}, but may prove to be a superior mass sensor than nanotubes due to its significantly larger surface area on which molecules can attach.

The performance of GNMR mass sensors is closely related to its quality (Q)-factor, which reflects the energy that is lost during the mechanical oscillation of the resonator.  In the first reported study of GNMRs, Bunch \emph{et al.} observed very low ($<100$) Q-factors for GNMRs working in the megahertz range~\cite{BunchJS}.  In a later experiment, the same group reported a dramatic increase of the Q-factor at lower temperatures, with the Q-factors reaching up to 9000 at 10K for GNMRs produced from the chemical vapor deposition growth method~\cite{Zande}.  Chen {\it et al.} also found that the Q-factor of GNMR increases with decreasing temperature, and reaches $10^{4}$ at 5 K~\cite{ChenC}.  More recently, Eichler {\it et al.}~\cite{EichlerA} found that the Q-factors of GNMRs can reach values of $10^{5}$, which is very close to the theoretical Q-factor upper bound of about $10^{6}$, which has been predicted for GNMRs with all edges fixed coupled with idealized mechanical actuation~\cite{KimSY,JiangJW2012}.

Because a key sensing objective for GNMRs is to enable detection of individual molecules or atoms, it is critical to determine methods of enhancing their mass sensitivity.  For example, Eichler {\it et al.} have shown that it is possible to increase the resonant frequency (and thus the mass sensitivity) of GNMRs by driving its mechanical oscillation into the nonlinear regime, which was explained using a continuum mechanics model~\cite{EichlerA}.  Using similar continuum models, several groups have theoretically studied the nonlinear effect on the mass sensitivity of different nanoresonators~\cite{BuksE,DaiMD,AtalayaJ}, with only a single, recent study related to graphene~\cite{eom2012}.  Besides the nonlinear resonance, mechanical strain is another possible method to improve the mass sensitivity of the GNMR, as the resonant frequency can be enhanced by mechanical tension~\cite{KimSY2008,KimSYapl,DaiMD,KimSYnanotechnology,QiZ}.

In this letter, we investigate the utility of inducing nonlinear oscillations to enhance the mass sensitivity of GBMRs using classical molecular dynamics (MD) simulation. Our simulations show that the adsorption-induced frequency shift resulting from a single Au atom can be enhanced by increasing the actuation energy, when the actuation energy parameter $\alpha$ is below a critical value $\alpha_{c}=2.0875\pm 0.0125$. For actuation energy above the critical value, the adsorbed Au atom exhibits significant diffusion, which reduces the Q-factor of the GNMR by more than three orders of magnitude and significantly degrades the mass sensitivity. Quite different from the widely used continuum mechanics explanation, we show that the nonlinear-induced frequency enhancement results from the `effective strain' that is induced in the oscillating GNMR. We derive an analytic expression for the effective strain $\epsilon_{\alpha}= \frac{3}{4}\pi^{2}\alpha\frac{E_{k}^{0}}{m\omega^{2}L^{2}}$ (here $E_{k}^{0}$ is the kinetic energy, $m$ the mass, $\omega$ the angular frequency, and $L$ length of the resonator) that enables us to directly link the equivalence of applied mechanical tensile strain and the strain induced by nonlinear oscillations to the resonant frequencies of the GNMRs.

\section{simulation details}

The graphene sample in our simulations has dimensions $(L_{x}, L_{y}) = (34, 38)$~{\AA}, and is composed of 504 carbon atoms.  The atoms at the $+x$ and $-x$ ends of the GNMR are fixed, while periodic boundary conditions are applied in the $y$ direction. The interactions of the carbon atoms are described by the Brenner (REBO-II) potential~\cite{Brenner}. For the cases where a single Au atom is adsorbed on the GNMR, the interaction between the Au atom and the GNMR is modeled by a Lennard-Jones potential with length parameter $\sigma$=2.9943~{\AA} and energy parameter $\epsilon$=0.02936~{eV}~\cite{KimSYnanotechnology}.  The standard Newton equations of motion are integrated in time using the velocity Verlet algorithm with a time step of 1 fs.

Our simulations are performed as follows.  First, the Nos\'e-Hoover\cite{Nose,Hoover} thermostat is applied to thermalize the system to a constant temperature of 4.2~{K} within the NVT ensemble, which is run for $10^{5}$ MD steps. The mechanical oscillation of the resonator is then actuated by adding a velocity distribution to the system, which follows the morphology of the first flexural vibrational mode of graphene~\cite{JiangJW2012}.  The imposed velocity distribution, or actuation energy, is $\Delta E=\alpha E_{k}^{0}$, where $E_{k}^{0}$ is the total kinetic energy in the GNMR after thermalization but just before its actuation and $\alpha$ is the actuation energy parameter.  After the actuation energy is applied, the system is allowed to oscillate freely within the NVE ensemble for $2^{19}$ MD steps. The data from the NVE ensemble is used to analyze the mechanical oscillation of the GNMR.

\section{results}

\begin{figure}[htpb]
  \begin{center}
    \scalebox{1.2}[1.2]{\includegraphics[width=8cm]{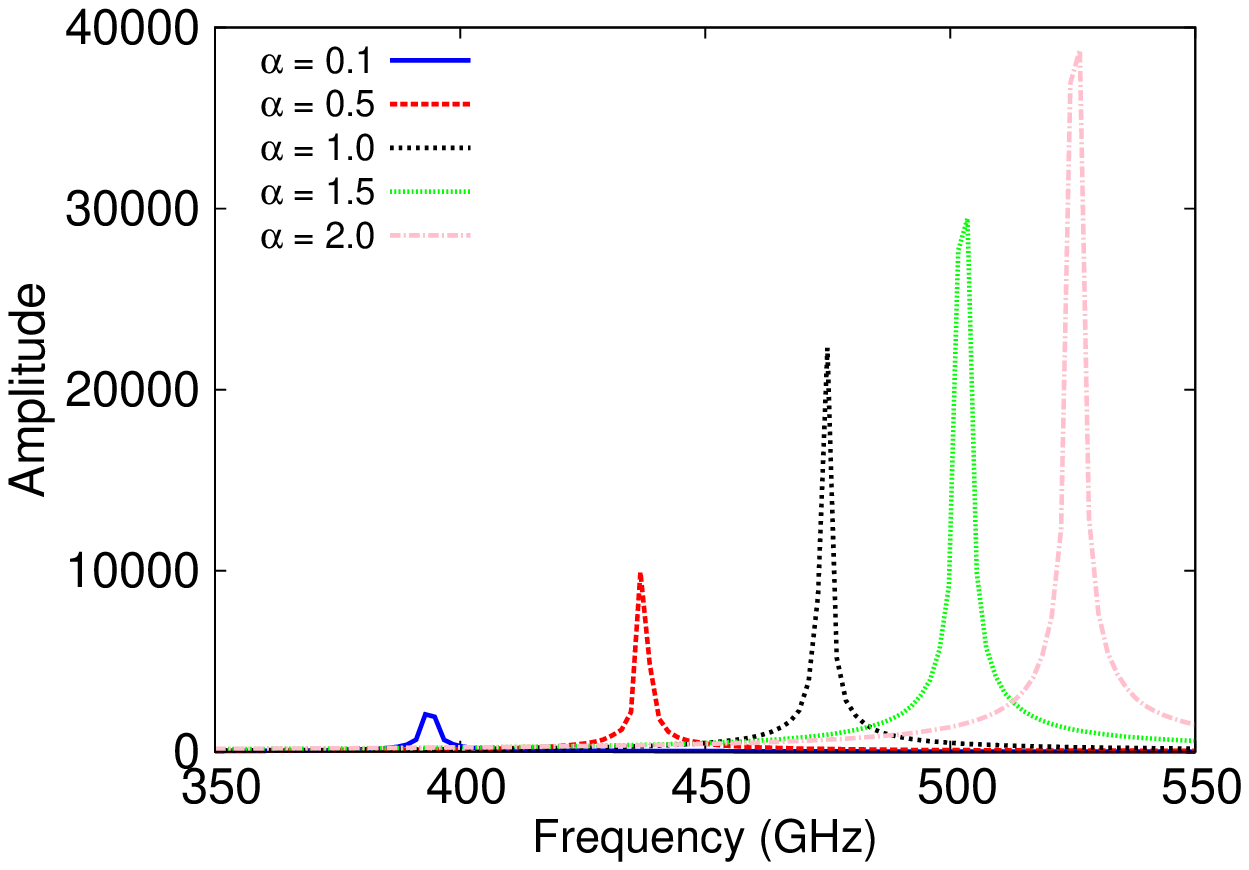}}
  \end{center}
  \caption{(Color online) Resonant frequencies for GNMRs. From left to right are pure GNMR at 4.2~K with different actuation energy parameters $\alpha=$ 0.1, 0.5, 1.0, 1.5, and 2.0.}
  \label{fig_fft_alpha}
\end{figure}

During the free vibration period, the energy in the GNMR switches between kinetic and potential energy. The frequency of the switching is $2f$, with $f$ being the frequency of mechanical oscillation.  Fig.~\ref{fig_fft_alpha} shows the resonant peaks that are obtained by taking a Fourier transformation of the time history of the kinetic energy; the resonant peaks are used to extract the resonant frequency $f$ at liquid helium temperature (4.2 K) for GNMRs with different actuation energies $\alpha=$ 0.1, 0.5, 1.0, 1.5, and 2.0, where for reference, $\alpha=1$ means that the actuation energy equals the total kinetic energy in the GNMR after thermalization. It should be noted that the amplitude in Fig.~\ref{fig_fft_alpha} is large, because of the long simulation time we have performed.  We focus on the liquid helium temperature because this temperature is commonly utilized in experiments involving GNMRs (eg. 90~{mK} or 4.0~{K} in Ref.~\onlinecite{EichlerA}).

\begin{figure}[htpb]
  \begin{center}
    \scalebox{1.2}[1.2]{\includegraphics[width=8cm]{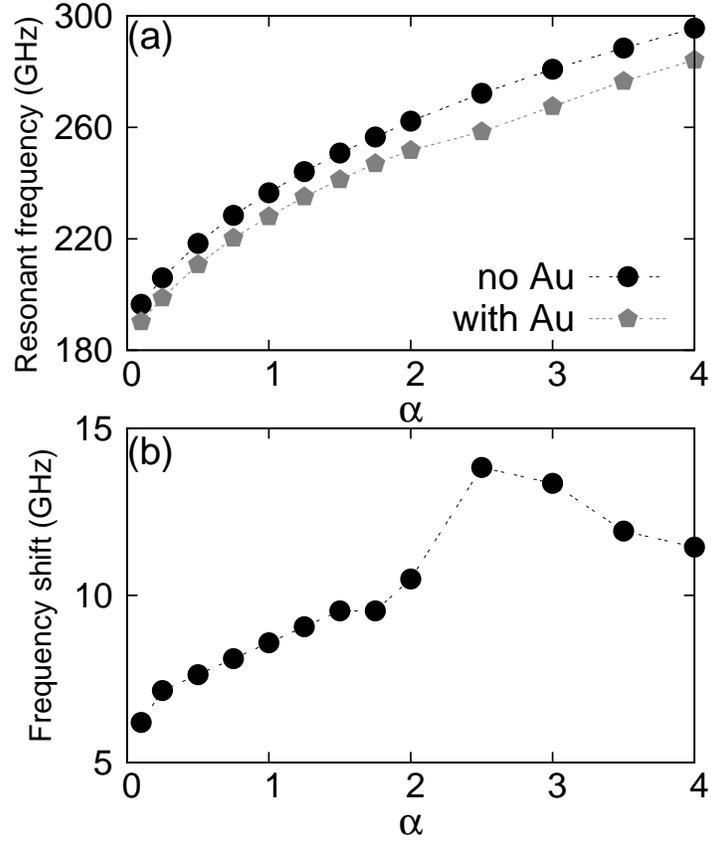}}
  \end{center}
  \caption{(Color online) The resonant frequency versus actuation energy parameter $\alpha$. (a) The resonant frequency for pure GNMR (circle points) and GNMR adsorbed by one Au atom (pentagonal points). (b). The magnitude of frequency shift induced by the adsorption of a single Au atom for different actuation energy parameters $\alpha$.}
  \label{fig_f_au_pure}
\end{figure}

Fig.~\ref{fig_f_au_pure} shows the resonant frequency as a function of actuation energy $\alpha$. Panel (a) shows that the resonant frequency is enhanced by increasing $\alpha$. This enhancement is due to the increasingly nonlinear behavior of the GNMR induced by the increasing oscillation amplitude.  Adsorption of a single Au atom on top of the GNMR causes a considerable reduction of the resonant frequency, due to the resulting increase in effective mass of the GNMR.  This frequency shift is what is measured experimentally to detect the adsorbed mass, and it is seen in Panel (b) that the magnitude of the frequency shift increases by a factor of three for large actuation energies ($\alpha\approx2.5$) as compared to if the GNMR was actuated with a small $\alpha$.  Panel (b) shows that the frequency shift can be increased by applying larger actuation energies, which also results in increased mass sensitivity.  However, as shown in the same figure, the frequency shift does not increase monotonically, and a decrease with increasing actuation energy is observed when $\alpha$ is large.

\begin{figure}[htpb]
  \begin{center}
    \scalebox{1.2}[1.2]{\includegraphics[width=8cm]{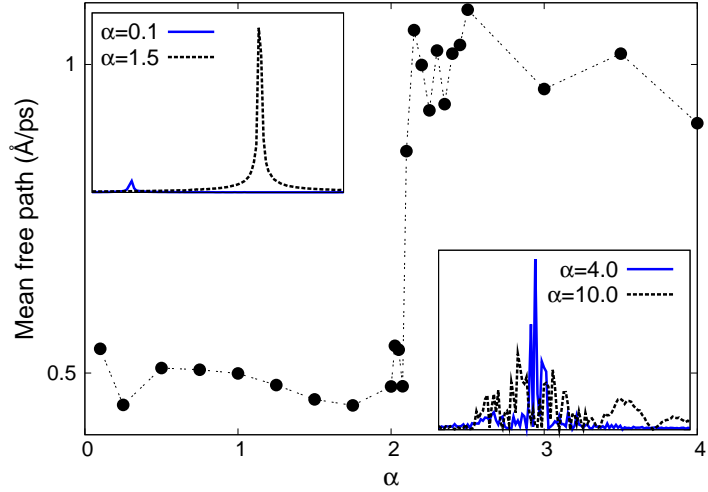}}
  \end{center}
  \caption{(Color online) The mean free path of the Au atom as a function of the actuation energy parameter $\alpha$. A sharp jump in the mean free path occurs at $\alpha_{c}=2.0875\pm 0.0125$, implying the diffusion of the Au atom for $\alpha>\alpha_{c}$. Left top inset shows smooth resonant curves for smaller actuation energy parameters $\alpha=$ 0.1 and 1.5. Right bottom inset shows significant diffusion-induced noise in the resonant curves for large actuation energy parameters $\alpha=$ 4.0 and 10.0.}
  \label{fig_meanfreepath}
\end{figure}

The reduction in the frequency shift with increasing $\alpha$ is determined to be a result of diffusion of the adsorbed Au atom, which was previously observed to occur at GNMR temperatures exceeding about 30~{K}~\cite{KimSYnanotechnology}.  To confirm the diffusion mechanism, we calculate in Fig.~\ref{fig_meanfreepath} the mean free path for the Au atom as a function of $\alpha$, where a sharp jump in the mean free path is observed at a critical value of actuation energy $\alpha_{c}=2.0875\pm0.0125$. For $\alpha<\alpha_{c}$, the mean free path of the Au atom is around 0.5~{\AA/ps} with small fluctuations.  In contrast, for $\alpha>\alpha_{c}$, the mean free path increases to a value around 1.0~{\AA/ps}, which implies diffusion of the Au atom.

\begin{figure}[htpb]
  \begin{center}
    \scalebox{1.0}[1.0]{\includegraphics[width=\textwidth]{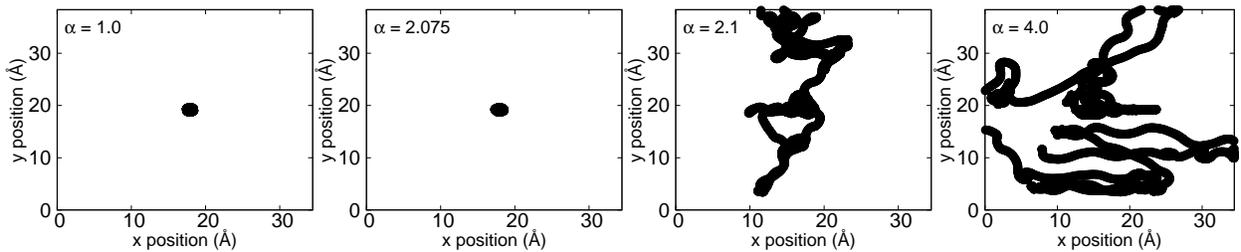}}
  \end{center}
  \caption{The trajectory history of the adsorbed Au atom for actuation energy $\alpha=$ 1.0, 2.075, 2.1, and 4.0 from left to right. The first two panels show a localized vibration of the Au atom, while the last two panels clearly show diffusion of the Au atom.}
  \label{fig_trajectory}
\end{figure}

\begin{figure}[htpb]
  \begin{center}
    \scalebox{1.2}[1.2]{\includegraphics[width=8cm]{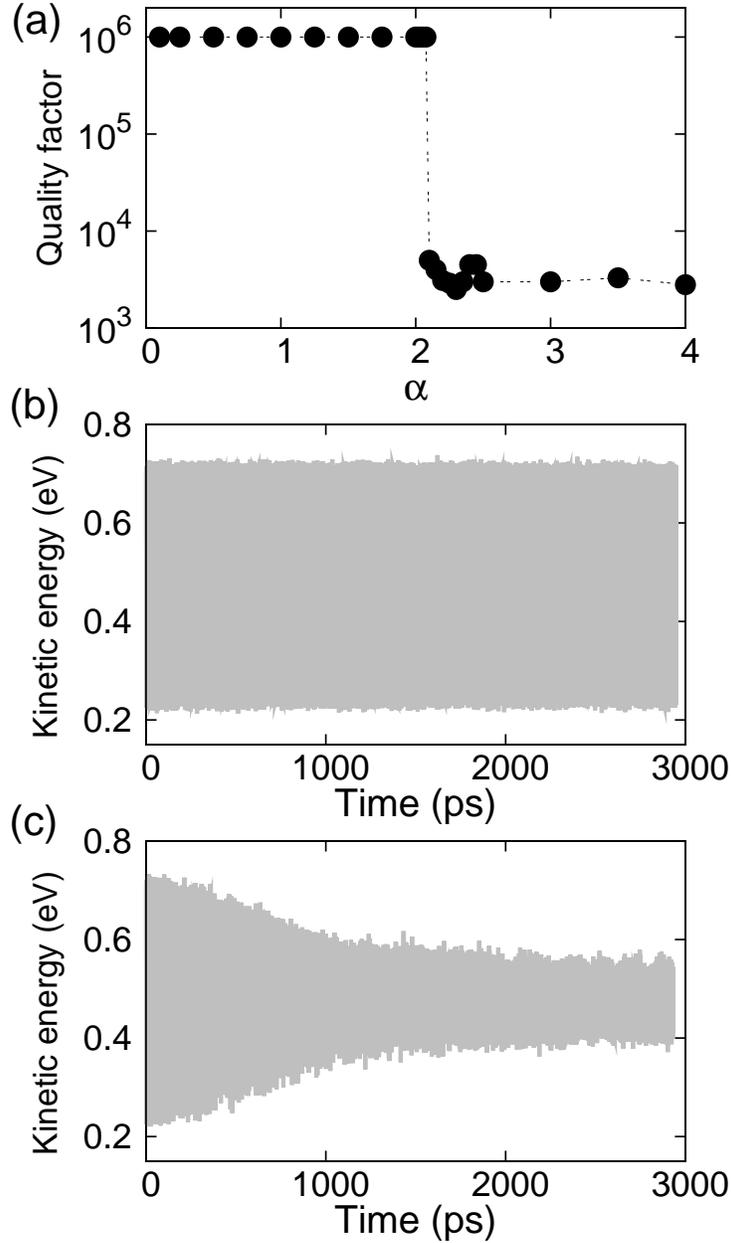}}
  \end{center}
  \caption{(Color online) The effect of the diffusion of the Au atom on the Q-factor of the GNMR. (a). The Q-factors of Au adsorbed GNMR remains extremely high value for $\alpha<\alpha_{c}$, but becomes three orders of magnitude smaller for $\alpha$ greater than the critical value $\alpha_{c}$. The time history of the total kinetic energy is shown in (b) for $\alpha=2.075$ and in (c) for $\alpha=2.1$, which are very close to $\alpha_{c}$. The energy decay in (c) is caused by the diffusion of the adsorbed Au atom shown in Fig.~\ref{fig_trajectory}.}
  \label{fig_quality_factor_alpha}
\end{figure}

The left top inset of Fig.~\ref{fig_meanfreepath} shows resonant curves for smaller actuation energy $\alpha=$ 0.1 and 1.5. These smooth curves provide evidence that the mechanical oscillation of the GNMR is the only vibrational mode in the system. The right bottom inset of Fig.~\ref{fig_meanfreepath} shows a significant amount of diffusion-induced noise in the resonant curves of large actuation energy $\alpha=$ 4.0 and 10.0. The noise corresponds to other vibrations that are induced by the diffusion of the Au atom.  Fig.~\ref{fig_trajectory} illustrates the trajectory history of the Au atom with $\alpha=$ 1.0, 2.075, 2.1, and 4.0. The diffusion is clearly observed for actuation energy above the critical value $\alpha_{c}$. The thermal noise provides important energy damping channels for the mechanical oscillation of the GNMR. As a result, the Q-factor of the GNMR is greatly reduced as shown in Fig.~\ref{fig_quality_factor_alpha}. There is almost no energy dissipation in GNMR for actuation energy below $\alpha_{c}$, leading to extremely high Q-factors. For actuation energy above the critical value, the Q-factor is reduced by more than three orders of magnitude. The time history of the kinetic energy is shown in Fig.~\ref{fig_quality_factor_alpha}(b) for $\alpha=2.075$ and in Fig.~\ref{fig_quality_factor_alpha}(c) for $\alpha=2.1$, from which the Q-factor has been extracted~\cite{JiangJW2012}.

\begin{figure}[htpb]
  \begin{center}
    \scalebox{1.5}[1.5]{\includegraphics[width=8cm]{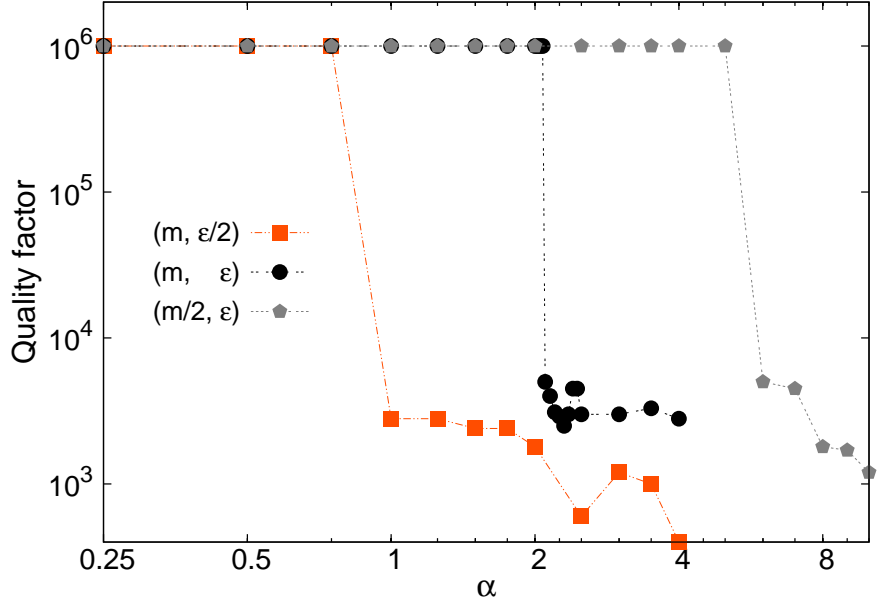}}
  \end{center}
  \caption{(Color online) Quality factors versus the actuation energy parameter $\alpha$ for graphene nanomechanical resonators adsorbed by a single atom with mass and Lennard-Jones potential parameter as ($m$, $\epsilon$), ($m/2$, $\epsilon$), and ($m$, $\epsilon/2$). }
  \label{fig_quality_factor_absorbate}
\end{figure}

Fig.~\ref{fig_quality_factor_absorbate} shows the Q-factor vs. the actuation energy parameter $\alpha$ for graphene nanomechanical resonators with three different adsorbates. In these three systems, the adsorbate mass $m$ and the Lennard-Jones interaction parameter $\epsilon$ are ($m$, $\epsilon$), ($m/2$, $\epsilon$), and ($m$, $\epsilon/2$). For the first system ($m$, $\epsilon$), the mass of the adsorbate is $m=197$ and the Lennard-Jones potential parameter is $\epsilon=0.02936$~{eV}, which are the actual values for Au adsorbate. For the second system ($m/2$, $\epsilon$), the mass of the adsorbate is reduced by half to be $m=99$ while $\epsilon$ (the interaction strength between the adsorbate and graphene) remains unchanged. In the third system ($m$, $\epsilon/2$), the mass of the adsorbate is unchanged at $m=197$ while the interaction strength $\epsilon$ between the adsorbate and graphene is reduced by half. 

In all three systems, there is a sharp decrease in the Q-factor at different values of actuation energy parameter $\alpha$, which results from the diffusion of the adsorbate.  These results also follow physical intuition.  For example, diffusion happens at smaller $\alpha$ for adsorbates that have a weaker bonding strength with graphene.  In contrast, diffusion happens at larger $\alpha$ for adsorbates with less mass, since those atoms have a smaller kinetic energy at the same temperature as larger mass adsorbates, and therefore more energy via the nonlinear actuation parameter $\alpha$ is needed to induce diffusion for the smaller mass atoms.  These results may serve as useful guidelines for experimentalists to verify our theoretical predictions, as different types of adsorbates (of different mass or/and different interaction strengths with graphene) are usually observed experimentally.

\section{discussion}

The above discussion has established that the mass sensitivity of the GNMR can be enhanced by driving the mechanical oscillations into the nonlinear regime using a larger actuation energy.  The remainder of this article will be devoted to explaining the mechanism that enables the nonlinear-induced resonant frequency, and thus mass sensitivity enhancement.

\begin{figure*}[htpb]
  \begin{center}
    \scalebox{1.0}[1.0]{\includegraphics[width=\textwidth]{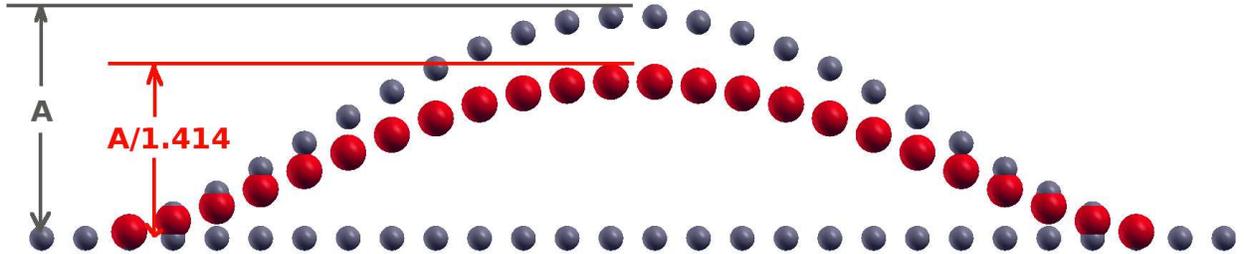}}
  \end{center}
  \caption{(Color online) Geometry of the GNMR. Before actuation, the GNMR lies in the horizontal plane as shown by the small gray points on the horizontal line. After mechanical actuation, the GNMR oscillates with amplitude $A$, which is determined by the actuation energy (see text for more details). The oscillating GNMR is regarded as an effective stationary shape with amplitude $A/\sqrt{2}$ (big red points). The difference between the length of the effective shape ($S$) and the length of the initial structure ($L$) yields an effective strain $\epsilon_{\alpha}=(S-L)/L$.}
  \label{fig_cfg}
\end{figure*}

As illustrated in Fig.~\ref{fig_cfg}, the GNMR is initially flat (gray points on the horizontal line) with length $L$.  After exciting the mechanical oscillations via the actuation energy $\Delta E=\alpha E_{k}^{0}$, the GNMR oscillates with amplitude $A=\sqrt{\frac{2\Delta E}{m\omega^{2}}}$, where $\omega=2\pi f$ is the angular frequency of the mechanical oscillation and $m$ is the total mass of the system. In the derivation of the oscillation amplitude, the thermal vibrations at 4.2~{K} have been ignored due to the very low temperature conditions.  We focus on a particular point in the GNMR, i.e. the atom at the midpoint, which oscillates as $u=A\sin\omega t$. The mean oscillation amplitude of this point is $\sqrt{<u^{2}>}=A/\sqrt{2}$, and so this point can effectively be regarded as a stationary point located at $A/\sqrt{2}$.  Similarly, the shape of the oscillating GNMR is equivalent to a stationary sine function, but with amplitude $A/\sqrt{2}$. This effective shape is shown as big red points in Fig.~\ref{fig_cfg}, where the length of the effective shape is
\begin{eqnarray}
S=\frac{\sqrt{2}L}{\pi}\sqrt{a^{2}+3}\left(K(k)\right),
\end{eqnarray}
where $a=\pi A/\sqrt{2}L$, $k=a^{2}/(a^{2}+1)$ and $K(k)$ is the complete elliptic integral of the first kind.

In analyzing the sine-shaped form of the oscillating GNMR (red dots) in Fig.~\ref{fig_cfg}, our key insight is that the oscillating GNMR is under an `effective strain' as compared to the original, flat GNMR (gray points on the horizontal line).  This effective strain is denoted by $\epsilon_{\alpha}$, where the subscript indicates that the effective strain is induced by the actuation energy $\alpha$, and where $\epsilon_{\alpha}=(S-L)/L$. Approximating the elliptic integral up to second order, we can derive an analytic expression for the oscillation-induced effective strain:
\begin{eqnarray}
\epsilon_{\alpha}= \frac{3}{4}\pi^{2}\alpha\frac{E_{k}^{0}}{m\omega^{2}L^{2}}.
\label{eq_effective_strain}
\end{eqnarray}
Substituting all structural and physical parameters into the expression, we get a concise relationship between the effective strain and the actuation energy that is applied to the GNMR: $\epsilon_{\alpha}\approx 0.216 \% \times \alpha$. The effective strain for the Au adsorbed GNMR is obtained analogously. As previously discussed, the resonant frequency of the GNMR can be enhanced by applying mechanical tension due to the enhanced stiffening of the structure that results~\cite{KimSY2008,KimSYapl,DaiMD,KimSYnanotechnology,QiZ}.  In this sense, we propose that the enhancement of the resonant frequency that results from driving the GNMR into the nonlinear oscillation regime is actually due to the effective strain that is induced in the oscillating GNMR.

\begin{figure}[htpb]
  \begin{center}
    \scalebox{1.0}[1.0]{\includegraphics[width=\textwidth]{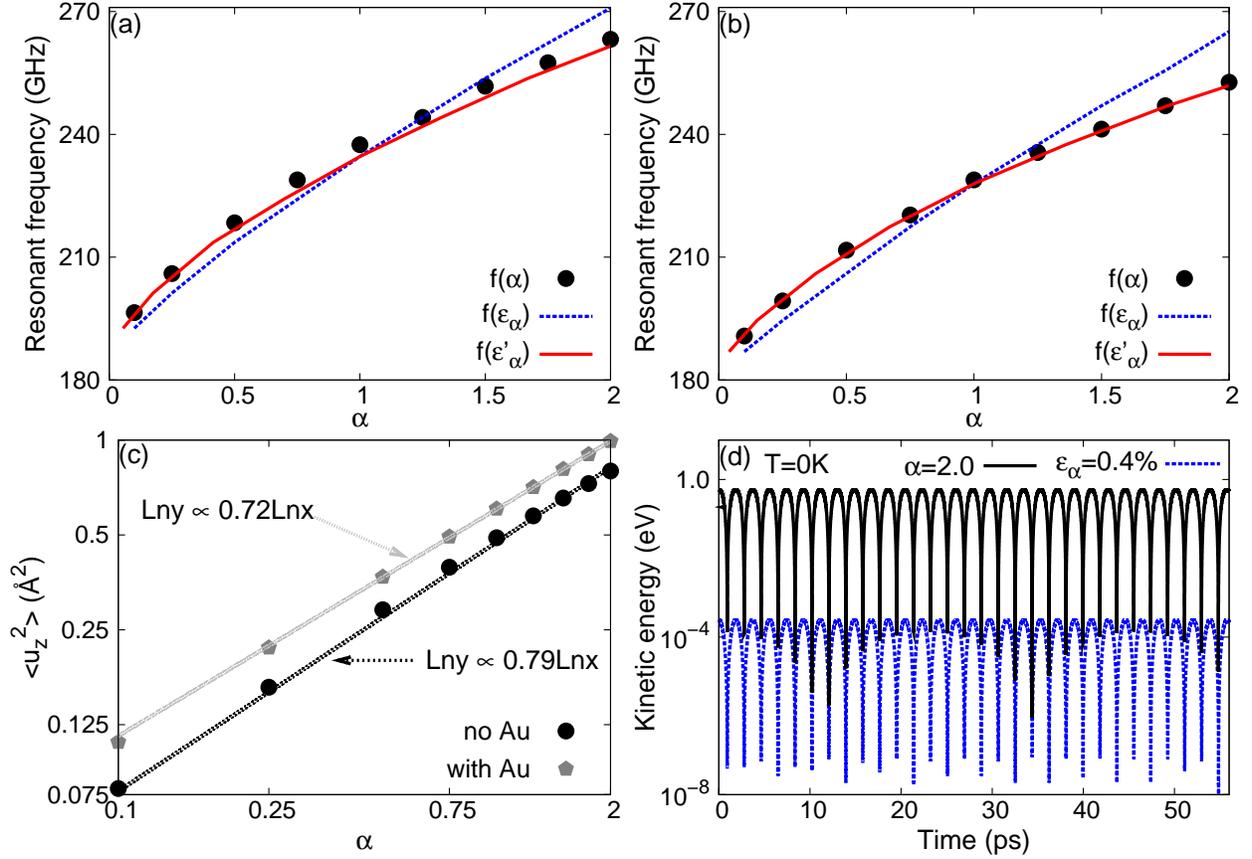}}
  \end{center}
  \caption{(Color online) Resonant frequencies for pure GNMR in (a) and GNMR with adsorption of an Au atom in (b). The resonant frequency can be enhanced by increasing actuation energy $\alpha$, i.e $f(\alpha)$ (circle points); or by applying effective strain, i.e $f(\epsilon_{\alpha})$ (dashed blue lines); or by applying a modified effective strain, i.e $f(\epsilon'_{\alpha})$ (solid red lines). (c) The square of the mean vibration amplitude for the atom in the center of the GNMR from the MD simulations; Lnx and Lny signify $\log_{e}x$ and $\log_{e}y$, respectively. (d) The time history of the kinetic energy for GNMR with actuation energy parameter $\alpha=2.0$ or for the GNMR under mechanical tension $\epsilon_{\alpha}=0.4\%$, both at 0~{K}.}
  \label{fig_f_alpha_strain}
\end{figure}

In Fig.~\ref{fig_f_alpha_strain}, we show that the effective strain is indeed responsible for the enhancement of the resonant frequency with increasing $\alpha$. We present the resonant frequency for the GNMR in (a) and the GNMR with adsorption of a single Au atom in (b). The circles represent the resonant frequency of the GNMR due to different actuation energies $\alpha$, i.e $f(\alpha)$, while the dashed lines (blue online) are the resonant frequencies $f(\epsilon_{\alpha})$ for the GNMR that was pre-stretched with a tensile strain $\epsilon_{\alpha}$ before actuation, where the strain $\epsilon_{\alpha}$ for actuation energy $\alpha$ is calculated by Eq.~(\ref{eq_effective_strain}). A very small actuation energy of $\alpha=0.001$ is used to actuate the stretched GNMR such only the strain effect is important in the GNMR under mechanical tension.

Both Figs.~\ref{fig_f_alpha_strain}(a) and (b) show good agreement between the resonant frequencies obtained from increasing the actuation energy $\alpha$ and by applying tensile strain $\epsilon_{\alpha}$. This agreement verifies that the effective strain is the cause for the increase of the resonant frequency that we have previously observed by increasing $\alpha$.  The agreement is particularly good for $\alpha<1.0$, though there is an increasing discrepancy for larger $\alpha$.
The solid lines (red online) show that this discrepancy is greatly improved by considering a high-order correction in the square of the oscillation amplitude $A^{2}$.  This high-order correction is demonstrated in Fig.~\ref{fig_f_alpha_strain}(c), where $u_{z}$ on the $y$-axis of Fig.~\ref{fig_f_alpha_strain}(c) is the oscillation of the midpoint of the GNMR obtained from the MD simulation, and where Fig.~\ref{fig_f_alpha_strain}(c) is plotted using a log-log scale.  $<u_{z}^{2}>$ deviates from linear behavior because of the phonon-phonon scattering. Although the temperature is quite low (4.2~{K}), the phonon-phonon scattering is still important for the first bending mode (i.e the mode of mechanical oscillation) because this mode is in a highly non-equilibrium state.  In other words, because the first bending mode has been mechanically actuated with very large amplitude, it is driven far away from its thermal equilibrium state at 4.2~{K}. Our MD simulations gives $<u_{z}^{2}>\propto \alpha^{0.79}$ for the pure GNMR. However, we have assumed $<u_{z}^{2}>\propto \alpha^{1.0}$ in our derivation of the effective strain. The effective strain $\epsilon_{\alpha}\propto <u_{z}^{2}>$, which is why we have $\epsilon_{\alpha} \propto \alpha$.  Therefore, the effective strain must be modified by: $\epsilon'_{\alpha}=\epsilon_{\alpha}/\alpha^{0.21}$, where $\epsilon_{\alpha}$ is the effective strain without correction.  A similar correction factor of $\alpha^{0.28}$ can be done for the effective strain of the GNMR with a single adsorbed Au atom.

We note that Fig.~\ref{fig_f_alpha_strain}(c) shows a small oscillation amplitude (~0.8~{\AA}) for the system simulated in this work, due to its small size. The amplitude $U$ actually depends on the length $L$ of the graphene as $U=2.43\times\frac{\sqrt{\alpha T}}{\omega(L)}$~{\AA}, where the frequency $\omega$ is length-dependent. For the system we have simulated, this formula gives $U\approx0.8$~{\AA} at $\alpha=1.0$, which is in good agreement with MD simulation results in Fig.~\ref{fig_f_alpha_strain}(c). Considering the flexural property of the bending mode in graphene, i.e $\omega\propto 1/L^{2}$, we can obtain the oscillation amplitude of an arbitrary system length as $U=U_{0}(L/L_{0})^{2}$, where $L_{0}=34$~{\AA} and $U_{0}=0.8$~{\AA} for our simulated system at $\alpha=1.0$. For instance, the mechanical oscillation amplitude becomes 67~nm if the system length is 100~nm. If the system length is 1~{$\mu$m}, then the mechanical oscillation amplitude is very large, i.e. 6.7~{$\mu$m}. These results show that $\alpha=1$ is large enough to induce oscillations that can be clearly distinguished from random thermal fluctuations in experiments.

Previously, we have established that the effective strain mechanism can explain the enhancement of the resonant frequency that occurs by increasing $\alpha$.  However, we also needed to perform a high-order correction for large actuation energies to account for the phonon-phonon scattering that is still present even at the low (4.2~{K}) temperature due to the highly non-equilibrium behavior resulting from the large actuation energy.  However, if the temperature is reduced to 0~{K}, there is no phonon-phonon scattering for the first bending mode because this mode cannot decay without the assistance of other vibrations according to symmetry selection rules~\cite{BornM}.  As a result, at 0~{K}, the enhancement of the resonant frequency by applying tensile strain before actuation should agree with that obtained by increasing the actuation energy, without any high-order correction.

Indeed, we observe this result in Fig.~\ref{fig_f_alpha_strain}~(d).  We emphasize that for these two simulations of the GNMR, the GNMR exists initially in an energy minimized configuration at 0~{K}, and the applied actuation energy is $\Delta E=\alpha E_{k}^{0}$.  Because the kinetic energy $E_{k}^{0}$ is 0 at 0~{K}, we use $E_{k}^{0}=E_{k}(4.2K)$, where $E_{k}(4.2K)$ is the total kinetic energy at 4.2~{K}, for both of the two simulations shown in Fig.~\ref{fig_f_alpha_strain}~(d). The solid line is the kinetic energy for the GNMR that is actuated by the large actuation energy parameter $\alpha=2.0$. The dashed line (blue online) is the kinetic energy of the GNMR that is pre-stretched in tension by $0.4\%$ before actuation, where $0.4\%$ is chosen as it is also the effective strain that results from the actuation energy $\alpha=2.0$. The pre-stretched GNMR is actuated with a very small actuation energy $\alpha=0.001$ so that the effect from $\alpha$ can be ignored and only strain is important in this case. We observe exactly the same resonant frequency in the mechanical oscillation in these two very different situations. From the Fourier transformation of the time history of thees two kinetic energy, we get the same resonant frequency of 270 GHz, which further establishes the equivalent effect of applied tensile strain and different actuation energies on the resonant frequencies of GNMRs.

\section{conclusion}

In conclusion, we used classical MD simulations to demonstrate that a significant enhancement of the mass sensitivity of graphene nanomechanical resonators can be achieved by driving them into the nonlinear oscillation regime.  The enabling mechanism was determined to be the effective strain that is induced in the nanoresonators due to the nonlinear oscillations.  A simple analytic expression relating the effective strain to the actuation energy was obtained, and shown to be quite accurate for moderate actuation energies.  The key implication is that it should be possible for experimentalists to directly incorporate the present findings to enhance the sensitivity of graphene-based mass sensors simply by actuating the graphene nanomechanical resonators into the nonlinear oscillation regime.

\textbf{Acknowledgements} The work is supported by the Grant Research Foundation (DFG) Germany.  HSP acknowledges support from the Mechanical Engineering Department of Boston University.

\pagebreak

\end{document}